\documentclass[journal=jacsat,manuscript=article]{achemso}
\setkeys{acs}{articletitle = true}
\usepackage{amsfonts,amsthm,amsmath,amssymb,bm}
\usepackage{multirow,tabularx,diagbox,slashbox}
\usepackage{url,hyperref}
\usepackage{graphicx}

\usepackage{tikz}

\title{Molecular Dynamics Simulations of Microscopic Structural Transition and Macroscopic Mechanical Properties of Magnetic Gels}

\author{Xuefeng Wei}
\affiliation{Wenzhou Institute, University of Chinese Academy of Sciences, Wenzhou, Zhejiang 325000, China}
\alsoaffiliation{School of Physical Sciences, University of Chinese Academy of Sciences, 19A Yuquan Road, Beijing 100049, China}
\author{Gaspard Junot}
\affiliation{Departament de F\'{i}sica de la Mat\`{e}ria Condensada, Universitat de Barcelona, 08028 Barcelona, Spain}
\author{Ramin Golestanian}
\affiliation{Max Planck Institute for Dynamics and Self-Organization (MPIDS), D-37077 G\"{o}ttingen, Germany}
\alsoaffiliation{Rudolf Peierls Centre for Theoretical Physics, University of Oxford, Oxford OX1 3PU, United Kingdom}

\author{Xin Zhou}
\alsoaffiliation{Wenzhou Institute, University of Chinese Academy of Sciences, Wenzhou, Zhejiang 325000, China}
\affiliation{School of Physical Sciences, University of Chinese Academy of Sciences, 19A Yuquan Road, Beijing 100049, China}

\author{Yanting Wang}
\affiliation{CAS Key Laboratory of Theoretical Physics, Institute of Theoretical Physics, Chinese Academy of Sciences, Beijing 100190, China}
\alsoaffiliation{Wenzhou Institute, University of Chinese Academy of Sciences, Wenzhou, Zhejiang 325000, China}
\alsoaffiliation{School of Physical Sciences, University of Chinese Academy of Sciences, 19A Yuquan Road, Beijing 100049, China}

\author{Pietro Tierno}
\affiliation{Departament de F\'{i}sica de la Mat\`{e}ria Condensada, Universitat de Barcelona, 08028 Barcelona, Spain}
\alsoaffiliation{Universitat de Barcelona Institute of Complex Systems, 08028 Barcelona, Spain}
\alsoaffiliation{Institut de Nanoci\`{e}ncia i Nanotecnologia, Universitat de Barcelona, 08028 Barcelona, Spain}
\author{Fanlong Meng}
\email{fanlong.meng@itp.ac.cn}
\affiliation{CAS Key Laboratory of Theoretical Physics, Institute of Theoretical Physics, Chinese Academy of Sciences, Beijing 100190, China}
\alsoaffiliation{Wenzhou Institute, University of Chinese Academy of Sciences, Wenzhou, Zhejiang 325000, China}
\alsoaffiliation{School of Physical Sciences, University of Chinese Academy of Sciences, 19A Yuquan Road, Beijing 100049, China}

\begin{document}

\begin{abstract}
Magnetic gels with embedded micro/nano-sized magnetic particles in crosslinked polymer networks can be actuated by external magnetic fields, with changes in their internal microscopic structures and macroscopic mechanical properties.
We investigate the responses of such magnetic gels to an external magnetic field, by means of coarse-grained molecular dynamics simulations.
We find that the dynamics of magnetic particles are determined by the interplay of between magnetic dipole-dipole interactions, polymer elasticity and thermal fluctuations. The corresponding microscopic structures formed by the magnetic particles such as elongated chains can be controlled by the external magnetic field.
Furthermore, the magnetic gels can exhibit reinforced macroscopic mechanical properties, where the elastic modulus increases algebraically with the magnetic moments of the particles in the form of $\propto(m-m_{\mathrm{c}})^{2}$ when magnetic chains are formed.
This simulation work can not only serve as a tool for studying the microscopic and the macroscopic responses of the magnetic gels,
but also facilitate future fabrications and practical controls of magnetic composites with desired physical properties.
\end{abstract}

\maketitle
\section{Introduction}
Magnetic gels, with micro/nano-sized magnetic particles embedded in crosslinked polymeric gels, belong to the class of smart magnetic materials, whose physical properties can be controlled by external magnetic fields~\cite{Brand2011,Thevenot2013,Weeber2018}.
By tuning such fields, e.g., the direction, the strength or the frequency, one can change the materials with different mechanical responses, such as material stiffness~\cite{Jolly1996,Collin2003,varga2005smart,varga2006magnetic} and deformations~\cite{raikher2000magnetodeformation,wang2011physical,Huang2015}.
Based on such field-controlling features, magnetic gels have been fabricated in laboratories and used as magnetic robots~\cite{Kim2018,Alapan2019,Kim2022}, sensors~\cite{gao2014magnetic,hankiewicz2016ferromagnetic,gloag2019advances}, actuators~\cite{shigetomi2020magnetic,he2023magnetic}, etc., and also for bio-medical applications~\cite{li2013magnetic,sung2021,veloso2021review} such as drug delivery/release due to the bio-friendly characteristics of magnetic fields.

Under an applied magnetic field, the materials can respond in different ways depending on how the magnetic particles are embedded in the polymer matrix.
For magnetic particles able to move freely in the polymer gel, they can form microscopic chains when the magnetic dipole-dipole interactions between particles dominate the thermal fluctuation as in ordinary solutions~\cite{Auernhammer2006,Weeber2013}, where the polymer gel serves mainly as a viscoelastic medium without much changes in its mechanical properties.
For superparamagnetic particles fixed in the polymer network~\cite{Junot2022}, the particles can still move but now they induce microscopic distortions in the polymer network, and they can even form magnetic chains if the magnetic dipole-dipole interactions are strong enough.
In the latter case, the responses of the embedded particles to the external magnetic field with simultaneous deformations of the polymer network underpins the physical mechanism for controlling the mechanical properties of the magnetic gels. 

\begin{figure}[htpb]
  \centering
  \includegraphics[width=0.78\columnwidth]{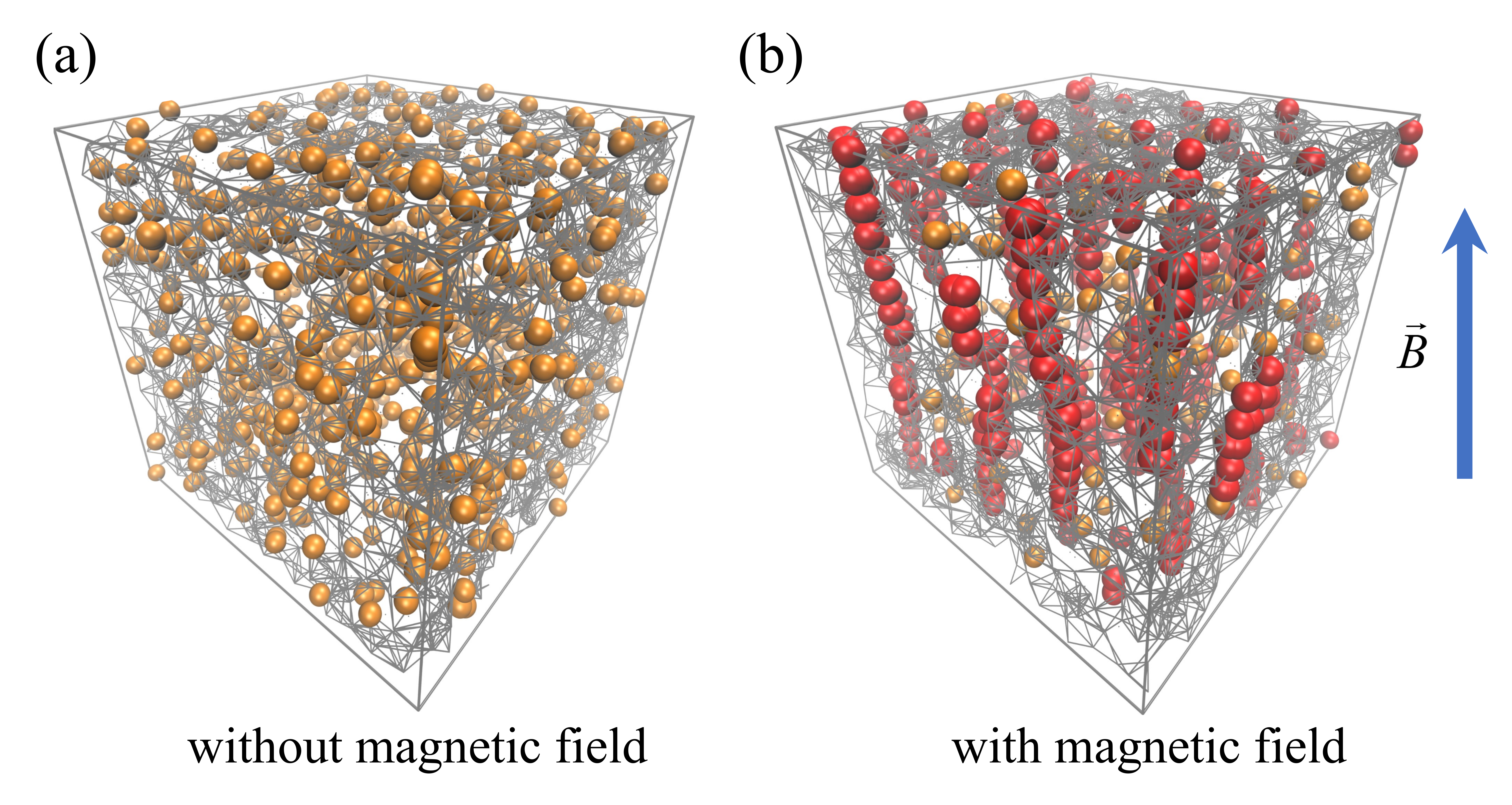}
  \caption{Magnetic gels with (a) uniformly distributed magnetic particles without the magnetic field and (b) magnetic chains under an external magnetic field. Spheres denote the magnetic particles and grey lines denote partial connections in the polymer network. }
  \label{fig1}
\end{figure}

The connection between the microscopic motion of the magnetic particles and the macroscopic responses of the material, is the key for understanding how external magnetic fields can be utilised to control the material properties.
Experimental methods such as small-angle X-ray scattering (SAXS)~\cite{sunaryono2016small}, small angle neutron scattering (SANS)~\cite{lawrence2018structure}, electron magnetic resonance (EMR)~\cite{ sorokina2012magnetic}, etc., are utilised for detecting the internal structure of the magnetic gels with macroscopic mechanical properties which are measurable by tensile testing~\cite{ramanujan2006mechanical,wu2011physically}, dynamic mechanical analysis~\cite{ponton2005regeneration,zhang2022influence}, etc.
Theoretical attempts include analytical modelling such as the lattice models~\cite{Ivaneyko2014,Pessot2016} and continuum models~\cite{Jarkova2003,Romeis2016,Junot2022}, and numerical methods such as finite element calculations~\cite{spieler2013xfem}, Monte Carlo simulations~\cite{Wood2011,Cremer2017} and molecular dynamics simulations~\cite{Weeber2012,Pessot2015,Weeber2015}.
In this work, we use coarse-grained molecular dynamics simulations to investigate the microscopic motion and structures of the magnetic particles, together with the macroscopic mechanical properties especially about the anisotropic elasticity of the materials.

\section{Simulation methodology}
\emph{Polymer network.} We follow Ref.42 to construct the 3D packing-derived (PD) polymer network. We first randomly place $N=W^3$ radially bidisperse spheres with harmonic repulsive interactions within a cubic unit cell of side length $W$ ($W=20$ is taken in later discussions), where half of the spheres are assigned with radius $r=r_0$ and the other half with $r= \nu r_0$ ($\nu = 1.4$ is chosen to avoid long-range crystal order~\cite{Shivers2019, dagois2012soft, koeze2016mapping}).
By continuously increasing the radius $r_0$ from $r_0=0$, the spheres will reach a jammed state at $r_0^{\mathrm{J}}$.
Then, from this disordered packing state, we generate a contact network by connecting the centers of the overlapping spheres (excluding rattlers) with springs at their rest lengths and angles. For sufficiently large systems, this procedure generates contact networks with the connectivity $z=2d$, where $d$ is the system dimensionality.
After generating the underlying network structure, we remove randomly chosen bonds and any consequent dangling ends repeatedly until the network reaches the desired average connectivity $z$.
The bonds connecting neighbouring nodes are modelled as Hookean springs and the bending effects is considered by introducing the bending energy.
The stretching and the bending energy of the network can be written as:
\begin{eqnarray}
U_{\mathrm{stretching}}(r)&=&\frac{\mu}{2}\sum_{i,j}\frac{(r_{ij}-r_{ij,0})^2}{r_{ij,0}}, \\  U_{\mathrm{bending}}(r)&=&\frac{\kappa}{2}\sum_{ijk}\frac{(\theta_{ijk}-\theta_{ijk,0})^2}{l_{ijk,0}},
\end{eqnarray}
respectively,
where $\mu$ and $\kappa$ denote the Hookean constant and the bending rigidity, respectively, $r_{ij,0}$ and $r_{ij}$ denote the length of spring $(ij)$ connecting the $i$th and the $j$th node before and after stretching, respectively,  $\theta_{ijk,0}$ and $\theta_{ijk}$ denote the angles between the spring $(ij)$ and $(jk)$ before and after bending, respectively,  and $l_{ijk,0}=(r_{ij,0}+r_{jk,0})/2$ denotes the average length of the two connecting springs.

\emph{WCA potential.} We choose polymer nodes randomly with the number density $n_0$ to be replaced with magnetic particles, i.e., the magnetic particles are uniformly distributed in the prepared state.
The diameters of the magnetic particles are set as $\sigma_\mathrm{m}=1.0\sigma$ and other non-magnetic nodes are set as $\sigma_{\mathrm{n}}=0.2\sigma$.
In our simulations, the truncated Weeks-Chandler-Andersen (WCA) potential~\cite{Weeks1971} is used to simulate the non-bonded and non-magnetic interactions between particles, as:
\[
U(r) =
\begin{cases}
    4\epsilon \left[ \left(\frac{\sigma_{ij}}{r}\right)^{12} -\left(\frac{\sigma_{ij}}{r}\right)^{6}\right] +C & \text{if } r \geq r_{\mathrm{cutoff}},  \\
    0 & \text{if } r < r_{\mathrm{cutoff}},
\end{cases}
\]
where $r_{\mathrm{cutoff}}=2^{1/6}\sigma_{ij}$ is the cutoff distance for given pairs, $\sigma_{ij}=(\sigma_i+\sigma_j)/2$ is the average diameter of particle $i$ and $j$, and $C$ is a constant ensuring the continuity of the potential energy at the cutoff distance.
In this work we take $\epsilon$ and $\sigma$ as the unit energy and length, respectively.

\emph{Magnetic dipole-dipole interactions.} The superparamagnetic particles can be magnetized under an external magnetic field $\bm{B}_{\mathrm{ext}}$, and they acquire magnetic moments $\bm{m} = \chi v_{\mathrm{p}}\bm{B}_{\mathrm{ext}}/\mu_0$ pointing along the direction of the magnetic field, where $\chi$ is the magnetic susceptibility, $v_p$ is the particle volume and $\mu_0$ is vacuum permeability.
These magnetic particles possess moments aligned along the field direction, and the interactions between the magnetic particles are:
\begin{align}
    U_{\mathrm{dd}}(\bm{r}_{ij})=\frac{ \mu_0 }{4 \pi}\left[\frac{1}{r_{ij}^3}(\bm{m}_i\cdot \bm{m}_j)-\frac{3}{r_{ij}^5}(\bm{m}_i\cdot \bm{r}_{ij})(\bm{m}_j\cdot \bm{r}_{ij})\right].
\end{align}
The magnetic dipole-dipole interactions between magnetic particles can be either attractive or repulsive, depending on their relative locations with respect to the separation $\bm{r}_{ij}$.

\emph{Molecular dynamics simulations.}
We perform the molecular dynamics simulations on a coarse-grained level based on the above model setup, using LAMMPS software~\cite{LAMMPS}.
The simulations are done in the canonical \emph{NVT} ensemble using a Langevin thermostat~\cite{Gronbech2013, Gronbech2020}.
The equation of motion for each particle is:
\begin{align}
    m \ddot{\bm{r}}=-\gamma \dot{\bm{r}} + \bm{F} + \bm{\xi},
\end{align}
where $m$ denotes the mass of the particle, $\gamma$ denotes the friction constant, $\bm{F}$ denotes the forces exerted by other particles and the polymer network, and $\bm{\xi}$ denotes the thermal noise due to the random collisions with implicit solvent molecules, which obeys fluctuation-dissipation theorem.
Periodic boundary conditions are adopted in the simulations and the detailed simulation procedure including simulation parameters can be found in the supplement~\cite{sup}.

\section{Results and discussions}

\subsection{Internal microscopic structures}
\begin{figure*}[htbp]  
  \centering
  \includegraphics[width=\textwidth]{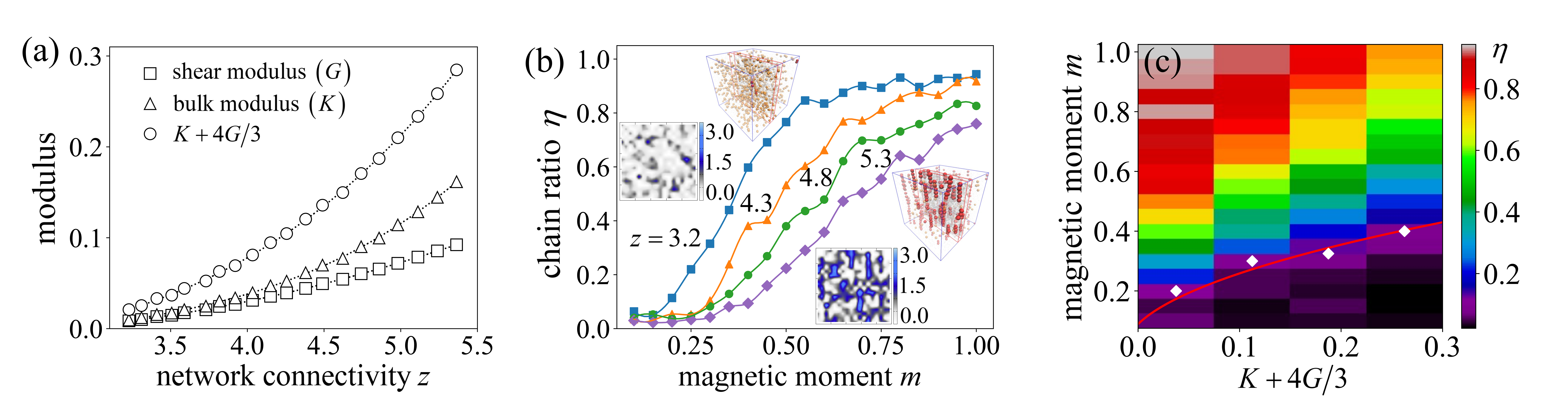}
  \caption{(a) Dependence of shear modulus $G$ (squares), bulk modulus $K$ (triangles) and $K+4G/3$ (circles) of the polymer gels on the network connectivity $z$. (b) Chain ratio as a function of the magnetic moments of the particles, in polymer networks with different connectivity. Insets denote the microscopic structures of the magnetic particles and the displacement field in the cross-sectional plane. (c) Chain ratio as a function of the magnetic moments and the network elasticity. Red line denote the physical conditions for the chain formation [Equation (5)], and the white diamonds correspond to simulation results of the critical chain ratio, $\eta=0.15$.}
  \label{fig2}
\end{figure*}

The elastic properties, e.g., the shear modulus and the bulk modulus, of the magnetic gels before applying any external magnetic field can be controlled by changing the connectivity in the polymer networks, as shown in Fig.~\ref{fig2}(a).
By increasing the connectivity of the polymer network, both the shear and the bulk modulus will increase, in an approximate form of $\propto (z-z_{\mathrm{c}})^{3}$ with $z_{\mathrm{c}}\simeq3.2$ as the critical connectivity for the polymer network to exhibit finite elasticity~\cite{Sharma2015,Shivers2019}.

As shown in Fig.~\ref{fig2}(b), the magnetic particles can form chain-like structures by increasing the strength of the magnetic field (magnetic moments of superparamagnetic particles), when the magnetic dipole-dipole interactions between the magnetic particles dominate the network elasticity and the thermal fluctuations;  
this phenomenon resembles the clustering instability found in magnetic microswimmers, where the microswimmers can form clusters when magnetic dipole-dipole interactions dominate thermal fluctuations~\cite{Meng2018,Meng2021}.
Due to the displacements of the magnetic particles in events of chain formations, the polymer network will be deformed with the local deformation field, as shown in the insets of Fig.~\ref{fig2}(b).
The chain ratio, $\eta$, defined as the ratio of the number of the magnetic particles forming the chains to the total number of the magnetic particles in the material, is introduced to characterize the amount of the micro-structures in the magnetic gel and also the relative strength of the magnetic dipole-dipole interactions over the network elasticity.
We use the following criterion to identify if a particle belong to a chain: the distance to its nearest neighboring magnetic particle is less than $2^{1/6}\sigma$, and the angle between the direction of the magnetic moment and the position vector connected to its nearest neighboring particle is less than $15^{\circ}$.
With the decrease in the network connectivity, i.e., the elastic moduli of the material, the chain ratio in magnetic gels of the particles with the same magnetic moments will increase.

By systematically changing the elastic moduli of the network and the magnetic moments of the particles, one can obtain the phase diagram showing the physical conditions for the chain formation, as shown in Fig.~\ref{fig2}(c).
In simulations, $\eta=0.15$ is taken as the value of the chain ratio, for denoting the conditions of the chain formation, and one can observe from Fig.~\ref{fig2}(c) that, for inclusions carrying large magnetic moments in soft polymer gels, they can easily form chain-like structures, coinciding with the theoretical conditions for the chain formation as derived in a previous work~\cite{Junot2022},
\begin{align}
    \frac{\mu_0 n_0 m^2}{k_BT} > \frac{v_{\mathrm{p}}}{k_BT}(K+4G/3)+1.
\end{align}

\begin{figure*}[htbp]
  \centering
  \includegraphics[width=0.78\textwidth]{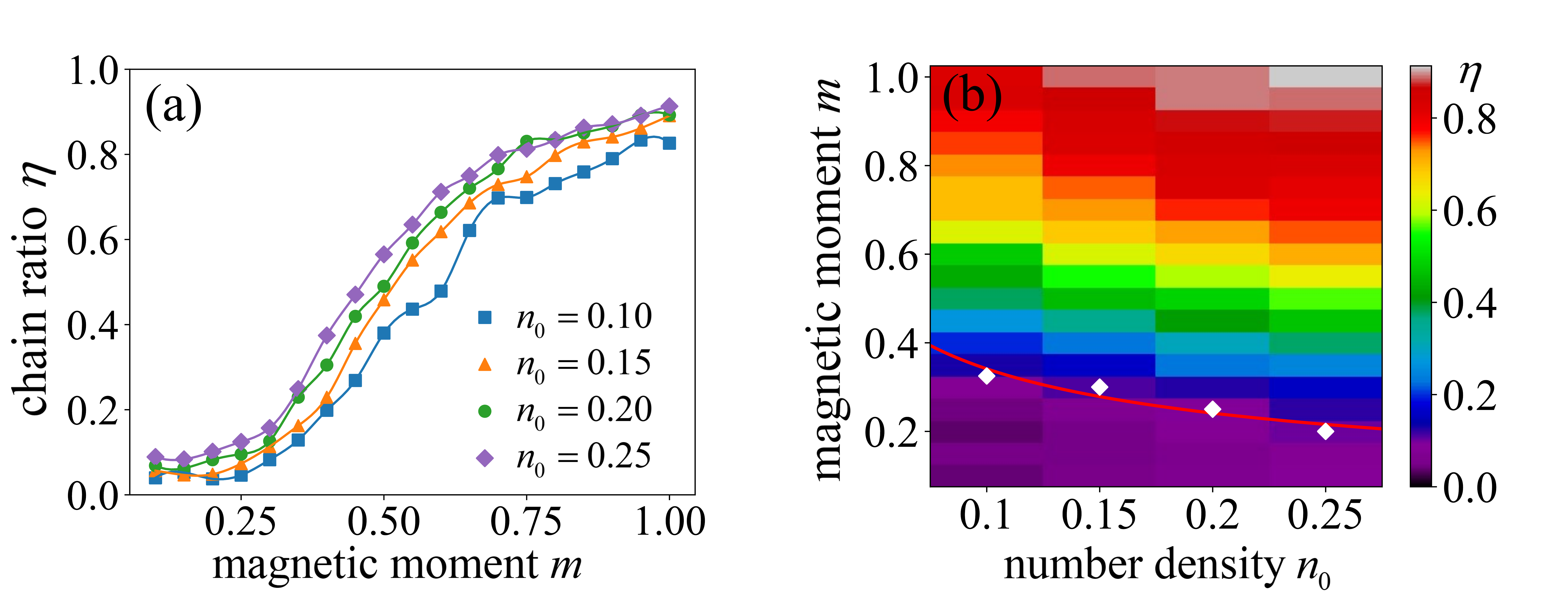}
  \caption{(a) Chain ratio $\eta$ as a function of the magnetic moments $m$ of the particles with different number density. (b) Dependence of the chain ratio on the magnetic moments and the number density of the magnetic particles. Red line denote the physical conditions for the chain formation [Equation (5)], and the white diamonds corresponds to simulation results of the critical chain ratio, $\eta=0.15$.}
  \label{fig3}
\end{figure*}

One can also check how the number density of magnetic particles in the prepared state, $n_0$, can influence the formation of the micro-structures in the gel.
As shown in Fig.~\ref{fig3}(a), the chain ratio is larger for systems characterized by a higher number density $n_0$, since the magnetic dipole-dipole interactions become stronger due to decreased distances between the magnetic particles.
The physical conditions for the chain formation in the plane of $(n_0, m)$ is provided in Fig.~\ref{fig3}(b), which also agrees with our theoretical prediction.

\subsection{Macroscopic mechanical properties}
The macroscopic mechanical properties of magnetic gels can become reinforced due to the formation of microscopic chains~\cite{Jolly1996,Collin2003,varga2005smart,varga2006magnetic}.
Here we investigate such effects by studying how magnetic gels consisting of particles carrying different magnetic moments can respond under shear deformations.

\begin{figure*}[htbp]
  \centering
  \includegraphics[width=\textwidth]{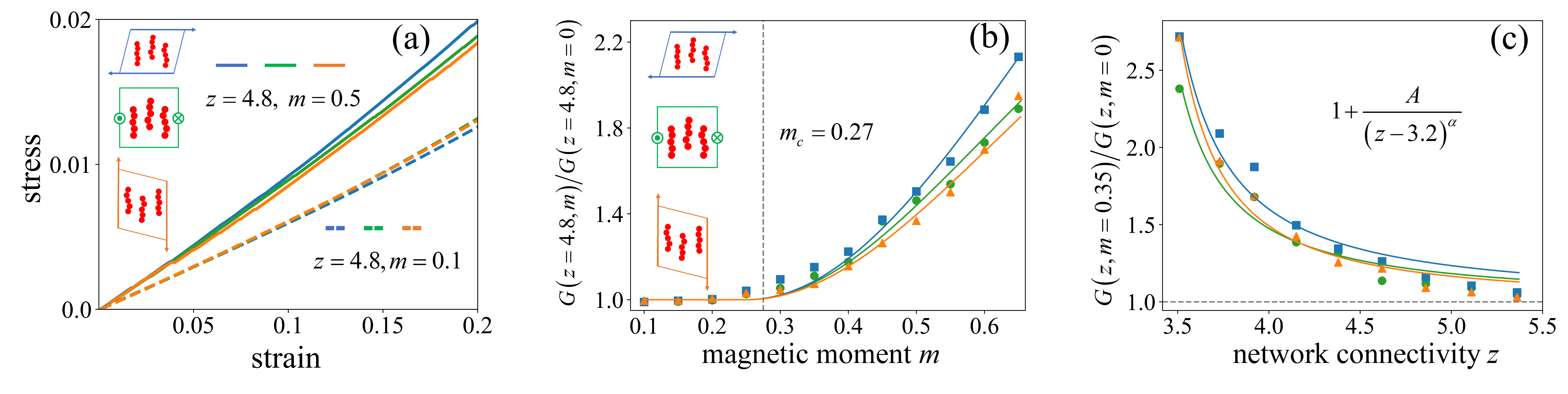}
  \caption{(a) Stress-strain relations of magnetic gels without (dash lines) and with (solid lines) chain formation, which undergoes the shear deformation in three orthogonal planes. (b) Dependence of the shear modulus $G_{\alpha\beta}$ as a function of the magnetic moment of the particles, and lines fitted with Equation~(\ref{G}). (c) Comparison between the magnetic interactions and the polymer elasticity in terms of the shear modulus of the magnetic gels. Lines fitted with the function $1+A/(z-z_{c})^{\alpha}$ with $A$ and $\alpha$ as fitting parameters. }
  \label{fig4}
\end{figure*}

As shown in Fig.~\ref{fig4}(a), the stress-strain relations of magnetic gels can change if the magnetic moments of the particles are different.
The magnetic gels with chain formation exhibit obvious mechanical reinforcements (larger stresses compared with those without chain formation at same strains) and weak anisotropy, whose mechanical responses depend on the applied plane of the shear deformation. 
By defining the shear modulus of the material as $G_{\alpha\beta}=\frac{d\sigma_{\alpha\beta}}{d\varepsilon_{\alpha\beta}}|_{\varepsilon_{\alpha\beta}=0}$, one can obtain the dependence of this modulus on the magnetic moments of the particles, as shown in Fig.~\ref{fig4}(b). There is a weak effect of the anisotropy in the shear moduli appears at $m_{c}\simeq 0.27$, corresponding to the chain ratio as $\eta=0.15$ (same criterion as used for chain formation). 
Compared with the material containing inherent chains in its relaxed state (springs not deformed)~\cite{Ivaneyko2014}, the anisotropy in the mechanical response here is weaker, since all the elastic springs are already distorted when the chains are formed in the current setup. 
More importantly, for $m\leq m_{c}$, the shear modulus remains almost unchanged by increasing the magnetic moments, but for $m> m_{c}$, the shear modulus increases quickly with the magnetic moments in an approximate form of
\begin{equation}\label{G}
  G_{\alpha\beta}(m)=G_{\alpha\beta,\infty}\frac{(m-m_c)^2}{a_m+(m-m_c)^2},
\end{equation}
where $G_{\alpha\beta,\infty}$, $a_m$ and $m_c$ are fitting parameters for specific materials~\cite{varga2005smart,Varga2006,Ivaneyko2014}.

By changing the network connectivity $z$, we can change the elasticity of the polymer gel in the prepared state, based on which one can compare the effects of magnetic interactions between the embedded particles and polymer elasticity in the mechanical properties of magnetic gels.
As shown in Fig.~\ref{fig4}(c), the contributions of magnetic interactions in the shear modulus of magnetic gel where magnetic chains are formed, are significant for $z\simeq z_{\mathrm{c}}$, which become less important by increasing the network connectivity due to the increased elasticity of the polymer network itself.

\section{Summary}
We have studied the microscopic structure formation and the macroscopic mechanical responses of magnetic gels, by utilising the  coarse-grained molecular dynamics simulations.
By increasing the strength of the magnetic dipole-dipole interactions between the particles, the microscopic chain structures can form, which induces reinforced mechanical responses of the magnetic gels.
This work can not only help to understand the connection between the microscopic structure and the macroscopic mechanical properties of the magnetic gels, but can also guide industrial fabrications of magnetic gels with desired mechanical responses and their controls by applying external magnetic fields for practical applications.

\bigskip

\section{Acknowledgments}
F. M. acknowledges supports by National Natural Science Foundation of China (Grant No. 12275332, 12047503, and 12247130), Chinese Academy of Sciences, Max Planck Society (Max Planck Partner Group), Wenzhou Institute (Grant No. WIUCASQD2023009) and Beijing National Laboratory for Condensed Matter Physics (2023BNLCMPKF005).
X. W. acknowledges supports from the China Postdoctoral Science Foundation (Grants No. 2023M743448) and the National Natural Science Foundation of China (Grants No. 12347170). P. T. acknowledges financial support from the European Research Council (Grant No. 811234) and from the Generalitat de Cataluña via the program “Icrea Academia”.
The computations of this work were conducted on the HPC cluster of ITP-CAS.

\providecommand{\latin}[1]{#1}
\makeatletter
\providecommand{\doi}
  {\begingroup\let\do\@makeother\dospecials
  \catcode`\{=1 \catcode`\}=2 \doi@aux}
\providecommand{\doi@aux}[1]{\endgroup\texttt{#1}}
\makeatother
\providecommand*\mcitethebibliography{\thebibliography}
\csname @ifundefined\endcsname{endmcitethebibliography}  {\let\endmcitethebibliography\endthebibliography}{}

\end{document}